
\magnification=\magstep1
\baselineskip=15truept
\font\tyt=cmcsc10 scaled\magstep4

\centerline{\tyt Penrose Inequality}
\medskip
\centerline{\tyt for Gravitational Waves}
\vskip 1.5cm
\centerline{\bf Janusz Karkowski\footnote{*}
{\rm e-mail: \tt Jkark@thrisc.if.uj.edu.pl}}
\centerline{\bf Piotr Koc\footnote{**}
{\rm e-mail: \tt ufkoc@ztc386a.if.uj.edu.pl}}
\vskip 0.5cm
\centerline{\bf Institute of Physics}
\centerline{\bf Jagellonian University}
\centerline{\bf Reymonta 4, 30-059 Krak\'ow,}
\centerline{\bf Poland}
\vskip 1.5cm
\centerline{\bf Zdobys\l aw \'Swierczy\'nski}
\vskip 0.5cm
\centerline{\bf Cracow Pedagogical University}
\centerline{\bf Institute of Physics}
\centerline{\bf Podchor\c a\.zych 2, Krak\'ow,}
\centerline{\bf Poland}
\vskip 3cm
\centerline{PACS number: 04.20.Cv}
\vskip 1cm
\centerline{\bf Abstract}
We investigate axially symmetric asymptotically flat vacuum self-gravitating
system. A class of initial data with apparent horizon was numerically
constructed. The examined solutions satisfy the Penrose inequality.
The prior analysis of a massive system and the present results suggest that
either massive or sourcefree configurations fulfill the Penrose inequality.
\vfill
\eject

\par
{\bf 1.Introduction.}
\par
The problem of the Cosmic Censorship Hypothesis is one of the most
significant issues of the classical gravity. The feasible
technique of testing it is the examination of the validity of the Penrose
inequality [1]. Breaking of the Penrose inequality points to the
violation of the Cosmic Censorship.
\par
The geometry created by massive nonspherical (spheroidal) bodies has been
numerically analyzed in [2]. The Penrose inequality in this case has been
satisfied. We would like to stress out that those calculations was
made under the assumption of conformal flatness.  Conformal flatness may be
intuitively understood as a restriction to the case without
gravitational waves - a conformally flat system without
matter is flat. Therefore the subject of our interest in this paper is in some
sense an orthogonal situation: a geometry of the collapsing pure gravitational
wave. After a few numerical tests [3] it has been recently analytically
established [4] that vacuum solutions with apparent horizon exist.
\par
In this paper we are numerically examining the following formulation of
the inequality:
\medskip
\par
\sl
Let $\Sigma$ be an asymptotically flat Cauchy surface with a non-negative
energy density, $m$ - the Arnowitt-Deser-Misner mass and $S$ the area of
an external apparent horizon. Then
$$m\geq\sqrt{S\over 16\pi}.\eqno(1)$$
\medskip
\rm
Penrose argues that if the above is not true then the Cosmic Censorship
will be broken. A more refined version of (1) has been proposed in [5].
The inequality (1) is still proven only for a restricted class
of geometries [6,7] (see also a short review in [8]).

\bigskip
\par
{\bf 2.Main equations.}
\par
We examine the formation of apparent horizons on the initial hypersurface
$\Sigma$ which is required to be {\bf momentarily static} and
{\bf axially symmetric} (Brill waves). In the case of vacuum constraints
on $\Sigma$ reduces to
$$^{(3)}R[g_{ij}]=0.$$
$^{(3)}R[g_{ij}]$ is 3-dimensional scalar curvature of $\Sigma$.
The axially symmetric line element may be written as follows [9]:
$$g_{ij}dx^idx^j=\Phi^4(r,\theta)\Bigl(e^{-q(r,\theta)}(dr^2+r^2d\theta^2)+r^2
\sin^2\theta d\varphi^2)\Bigr).$$
Due to vanishing of the scalar curvature, $\Phi$ and $q$ are constrained
by [9]
$$\Delta\Phi+f\Phi=0,\eqno(2)$$
where $\Delta$ denotes 3-dimensional flat laplacian and
$$f=-{1\over 8}\biggl({\partial^2 q\over\partial r^2}+
{1\over r}{\partial q\over \partial r}+
{1\over r^2}{\partial^2 q\over\partial \theta^2}\biggr).
\eqno(3)$$
Regularity of metric on the axis implies the boundary conditions for $\Phi$ and
$q$.
$${\partial\Phi\over \partial\theta}(\theta =0)=0,\qquad
{\partial q\over\partial\theta}(\theta =0)=0,\qquad q(\theta =0)=0.\eqno(4)$$
The equivalent group of conditions must hold for $\theta=\pi$.
Moreover asymptotically solution is flat: $\Phi\rightarrow 1+{m\over 2r}$,
$q\rightarrow 0$. Where $m$ is the ADM mass (see equation (1)).
The above requirements may be satisfied by following choices of function
$q$ [3]:
$$q_1(r,\theta)={Ar^2\sin^2\theta\over 1+r^5}.\eqno(5)$$
Inserting $q$ in (3) we obtain
$$f_1=-{A\over 8}\bigl(2U-45U^2r^5\sin^2\theta+50U^3r^{10}\sin^2\theta\bigr),
\eqno(6)$$
where
$$U={1\over 1+r^5}.\eqno(7)$$
Similarly we analyze another $q$:
$$q_2(r,\theta)=Ae^{-\alpha r^2}r^2\sin^2\theta.\eqno(8)$$
Now $f$ has the following form:
$$f_2=-{A\over 4}e^{-\alpha r^2}(1-6\alpha r^2\sin^2\theta +
2\alpha^2r^4\sin^2\theta).\eqno(9)$$
The above formulas we use in the next analysis for calculation of function
$\Phi$. Because of the regularity the determinant of the metric must have no
zeros. Therefore we consider only solutions which are everywhere positive;
$\Phi>0$.
\par
A surface $S$ is called {\bf an apparent horizon} if
the expansion of outgoing future directed
null geodesics which are orthogonal to $S$ vanishes everywhere on $S$.
In the momentarily static case expansion is the divergence
of normal unit vector $n^i$ to $S$.
$$D_in^i=0.\eqno(10)$$
Hence on time symmetric $\Sigma$ notion of apparent horizon coincides
with definition of minimal surface. We investigate the {\bf external apparent
horizon} which is the outermost apparent horizon that surrounds the region
of concentration of gravitational energy. Equation (10) for our metric
takes the form
$$ r_{\vartheta\vartheta}+{r_\vartheta^3\over r^2}
\biggl({4\Phi_\vartheta\over\Phi}-{q_\vartheta\over 2}+cot\vartheta\biggr)
-r_\vartheta^2\biggl({4\Phi_r\over\Phi}-{q_r\over 2}+{3\over r}\biggr)+$$
$$+r_\vartheta\biggl({4\Phi_\vartheta\over\Phi}-{q_\vartheta\over 2}+
cot\vartheta\biggr)-
r^2\biggl({4\Phi_r\over\Phi}-{q_r\over 2}+{2\over r}\biggr)=0\eqno(11)$$
The area of $S$ is given by
$$S=2\pi\int\limits_0^\pi\Phi^4e^{-q/2}\sqrt{r_\vartheta^2+r^2}
{}~~r\sin\vartheta d\vartheta\eqno(12)$$

\bigskip
\par
{\bf 3.Numerical techniques and results of calculations.}
\par
Originally we intended to solve the eq. (2) by the use one of standard
method on lattice. The numerical error was too big and therefore we solve
the partial differential equation (2) by
expanding the functions $\Phi$ and $f$ in a series of the Legendre polynomials
[3]. We obtained an infinite system of ordinary differential equations
for $r$ dependent functions. We approximately solved this system by
retaining only leading  terms (up to l=5). Numerical results shows that
the contribution from the higher terms is very small and safely can be
neglected. Our numerical analysis of the system just described and the
equation for external apparent horizon (11) was based of the fourth order
Runge-Kutta method. For $q_2$ we examined only one value of $\alpha=0.1$.
The solutions for other values of $\alpha$ may be obtained from the previous
ones by simple scaling transformations.
\par
We analyzed only such configurations for which the shape of the external
apparent horizon significantly diverges from spherical symmetry. In the
case of spherical symmetry the inequality (1) reduces to an identity.
In the below tables we collect the most important results of our calculations.
The first table corresponds to $q$ given by formula (5) and the
second one to (8) with $\alpha=0.1$.

\par
$$\vbox{\tabskip=0pt \offinterlineskip
\halign to380pt{\strut#& \vrule#\tabskip=1em plus2em&
\hfil#& \vrule#&\hfil#\hfil& \vrule#&\hfil#\hfil&
\vrule#&\hfil#\hfil& \vrule#&\hfil#\hfil&
\vrule#& \hfil#& \vrule#\tabskip=0pt\cr \noalign{\hrule}
& &\omit\hidewidth $A$\hidewidth& & \omit\hidewidth $m$\hidewidth
& &\omit\hidewidth $r_0$\hidewidth& & \omit\hidewidth $r_{\pi/2}$\hidewidth
& &\omit\hidewidth $S$\hidewidth
& &\omit\hidewidth $m-\sqrt{S/16\pi}$\hidewidth& \cr \noalign{\hrule}
& & 8   & &  4.59719 & & 2.13920 & & 2.08129 & & 1055.72607   &
& 0.014     &\cr \noalign{\hrule}
& & 8.5 & &  7.30543 & & 3.63536 & & 3.59755 & & 2681.55505   &
& 0.0015    &\cr \noalign{\hrule}
& & 9   & & 14.00554 & & 7.00300 & & 6.98652 & & 9858.96851   &
& 0.0006   &\cr \noalign{\hrule}
}}$$
\par
\centerline{table 1.}

\par
$$\vbox{\tabskip=0pt \offinterlineskip
\halign to380pt{\strut#& \vrule#\tabskip=1em plus2em&
\hfil#& \vrule#&\hfil#\hfil& \vrule#&\hfil#\hfil&
\vrule#&\hfil#\hfil& \vrule#&\hfil#\hfil&
\vrule#& \hfil#& \vrule#\tabskip=0pt\cr \noalign{\hrule}
& &\omit\hidewidth $A$\hidewidth& & \omit\hidewidth $m$\hidewidth
& &\omit\hidewidth $r_0$\hidewidth& & \omit\hidewidth $r_{\pi/2}$\hidewidth
& &\omit\hidewidth $S$\hidewidth
& &\omit\hidewidth $m-\sqrt{S/16\pi}$\hidewidth& \cr \noalign{\hrule}
& & -2.4 & & 14.76569 & & 5.13853 & & 7.90747 & & 10905.86482 &
& 0.0359 &\cr \noalign{\hrule}
& & -2.5 & & 16.63314 & & 6.47847 & & 8.85901 & & 13864.95839 &
& 0.0249 &\cr \noalign{\hrule}
& & -2.6 & & 18.77306 & & 7.85006 & & 9.86873 & & 17680.94401 &
& 0.0180 &\cr \noalign{\hrule}
}}$$
\par
\centerline{table 2}
\par
Summarizing, the obtained result confirm the validity of the
Penrose inequality for the analyzed class of Brill waves.
Although the proof of the inequality (1) is still uncompleted
it seems that our belief in this conjecture is rather reasonable.

\bigskip
\par
{\bf Acknowledgment:} This work is partly supported (PK) by the Polish
Government Grant no {\bf PB 2526/2/93} and {\it Stefan Wajs} foundation.

\eject
\par
\centerline{\bf References}
\par
[1] R.Penrose, Ann N Y Acad Sci 224, 125 (1973).\par
[2] J.Karkowski, E.Malec, Z.\'Swierczy\'nski, Class. Quantum Grav.
(1993) to be\break published.\par
[3] K.Eppley, Phys. Rev. D16, 1609 (1977).\par
[4] R.Beig and N.\'O Murchadha, Phys. Rev. Lett. 66, 2421 (1991).\par
[5] G.Horowitz, Springer Lecture Notes in Physics 202, 1 (1984).\par
[6] M.Ludvigsen, J.Vickers, J. Phys. A16, 3349 (1983).\par
[7] J.Jezierski, Class. Quantum Grav. 6, 1535 (1988).\par
[8] E.Malec, Acta Phys. Pol. B22, 829 (1991).\par
[9] D.Brill, Ann. Phys. 7, 466 (1959).\par

\end